\documentclass{article}
\usepackage{amsmath,amsbsy,amsfonts,amssymb}
\usepackage{subfigure}
\usepackage{graphicx}
\usepackage{hyperref}

\usepackage[left=1.5cm,right=1.5cm,top=2cm,bottom=2cm]{geometry}

\newcommand{\gfi}{\nabla_\text{T}\varphi}

\newcommand{\gofi}{\nabla_\text{T}^\perp\varphi}

\newcommand{\ktc} { {k_\text{T}}^2 }

\newcommand{\cte} {c_\text{TE} }
\newcommand{\ctm} {c_\text{TM} }
\newcommand{\ctectm} {\left( c_\text{TE}\,c_\text{TM}^*+c_\text{TE}^*\,c_\text{TM}\right) }

\author{Irving Rond\'on and F.~ Soto-Eguibar,\\
 Instituto Nacional de Astrof\'isica \'Optica y Electr\'onica,\\ Puebla, C.P. 72840 M\'exico.
}

\title{Electromagnetic field theory  for  invariant beams using scalar potentials}
\begin{document}

\maketitle
\begin{abstract}
	We present a description of the electromagnetic field for propagation invariant beams using scalar potentials. Fundamental dynamical quantities are obtained: the energy density, the Poynting vector and  the Maxwell stress tensor. As an example, all these quantities are explicitly calculated for the Bessel beams, which are the invariant beams  with circular cylindrical symmetry.
\end{abstract}

\section{Introduction}
Propagation invariant beams, also known as \textquotedblleft non-diffracting beams\textquotedblright,  propagate indefinitely without changing their transverse intensity distribution. These optical fields are the well known plane waves, the Bessel beams \cite{Durnin}, the Mathieu beams \cite{Julio} and the Weber beams \cite{Bandres}. These invariant beams correspond to the solutions of the Helmholtz wave equation in Cartesian, circular cylindrical, parabolic cylindrical, and elliptical cylindrical coordinates, respectively. These  non-diffracting beams have a large  quantity of applications in fundamental and  applied  science in areas such as quantum mechanics \cite{RJ,BM1,BM2,BM3}, acousto-optics \cite{Zhang},  non-linear optics \cite{Wulle}, optical tweezers \cite{Hugo1}, fluid dynamics \cite{Ivan},  and optical communications  \cite{Willner}, among others. \\
In order to obtain the maximum possible information from these fields, it is very important to have a deeper insight of their physical properties. Although the vector approach is the most common method used in electromagnetic theory, several interesting works have used a scalar formalism \cite{Durnin}, such as  \cite{Karem1} and \cite{Karem2}. We use the scalar approach in order to  obtain  the fundamental  dynamic quantities of those beams; dynamic quantities are present in several physical phenomena which occur in different contexts and scenarios, but are nevertheless governed by the same physical laws. To the best of our knowledge, the complete general explicit expressions have not been presented before; as a matter of fact, the Maxwell stress tensor has been scarcely mentioned in the previously published literature. Additionally, by considering the interference between modes, our results can give new insights.\\
It must be emphasized that the scalar potential approach and the more general vector approach are completely equivalent, in the understanding that the results predicted for the measured quantities are exactly the same in both approaches; however, in several circumstances, like the one presented here, the scalar approach is easier. The link between these two approaches is the Hertz potential formalism \cite{Stratton}, but a review of this is beyond the scope of this paper.\\
The article is organized as follows: In Section \ref{SectionScalarPotential}, starting from the Maxwell equations, we present the scalar potential approach. Then, an exact explicit expression, in terms of the scalar potential, is derived for the energy density of the invariant fields in Section \ref{SectionEnergyDensity}, and the same is done for the Poynting vector in Section \ref{SectionVectorPoynting}. In Section \ref{SectionStressTensor} the Maxwell stress tensor is calculated. In order to present concrete examples, we have evaluated the corresponding quantities for the Bessel beams in each section. To show the  usefulnesses of the method, in Section \ref{cilindro}, the force exerted by a zero order Bessel beam on a small cylinder is calculated using the Maxwell stress tensor. Finally, our conclusions are presented in Section \ref{conclusiones}.\\

\section{Scalar potentials}\label{SectionScalarPotential}
Any given electromagnetic field answers to the Maxwell equations, which in the International System of Units (SI) \cite{Stratton,Jackson,Griffiths} are
\begin{subequations}
	\begin{align}
	\nabla \cdot \vec{D} =&   \rho_f, \\ 
	\nabla \cdot \vec{B} =& 0,   \\ 
	\nabla \times  \vec{E}=& - \frac{\partial \vec{B}}{\partial t}, \\ 
	\nabla\times\vec{H} =&   \vec{J}_f +  \frac{\partial \vec{D}}{\partial t} ,	
	\end{align}
\end{subequations}
where the microscopic electric and magnetic fields are $\vec{E}$ and $\vec{B}$, the corresponding macroscopic fields are $\vec{D}$ and $\vec{H}$, $\rho_f$ is the free charge density, and $\vec{J}_f$ is the free current density. \\
The constitutive relations between micro and macroscopic vector fields are
\begin{subequations}
	\begin{align}
	\label{eqMaxConstitutive1}
	\vec{D}&=\varepsilon _0 \vec{E}+ \vec{P}, \\
	\label{eqMaxConstitutive2}
	\vec{B}&=\mu_0 \left( \vec{H}+ \vec{M}\right)  ,
	\end{align}
\end{subequations}
where the electric polarization, $\vec{P}$, is the average electric dipole moment per unit volume  and the magnetization, $\vec{M}$, is the average magnetic dipole moment per unit volume; the free space electric permittivity is $\varepsilon_0$ and $\mu_0$ is the free space magnetic permeability.\\
For the sake of simplicity, we will consider an isotropic linear homogeneous medium without losses, that means that the electric permittivity $\varepsilon$ and the magnetic susceptibility $\mu$ are both real constants; hence the constitutive relations reduce to $\vec{D} =\varepsilon \vec{E}$  and $\vec{B} =\mu \vec{H}$. As we are interested only in the propagation of the electromagnetic fields and not in its production, we will also suppose that there is no free charge density nor free currents. Without loss of generality, we will also suppose that all the fields are monochromatic, of frequency $\omega$. With all these considerations, we can rewrite the Maxwell equations as
\begin{subequations} 
	\begin{align}
	\nabla \cdot \vec{E} &= 0,  \\ 
	\nabla \cdot \vec{H} &= 0, \\
	\nabla \times  \vec{E} &= - i \omega \mu  \vec{H},  \label{eq:CurlE}\\
	\nabla\times\vec{H} &= i \omega \varepsilon  \vec{E}.  \label{eq:CurlH}
	\end{align}
\end{subequations}
If we take the curl of equations (\ref{eq:CurlE}) and (\ref{eq:CurlH}), and combine with the other two Maxwell equations, we obtain the Helmholtz vector equations for the electric field and for the magnetic field,
\begin{subequations}\label{eq:HelmholtzH}
	\begin{align}
	\nabla^2 \vec{E}   +  k^2 \vec{E} = 0, \label{eq:HelmholtzE}  \\
	\nabla^2 \vec{H}  +  k^2 \vec{H} = 0, 
	\end{align}
\end{subequations}
where we have defined the wave-vector magnitude $k^2 = \left( \omega/v \right)^2$ in terms of the speed of light in the medium, $v^2= \dfrac{1}{\mu \varepsilon}$.\\

Let us write the electromagnetic fields as \cite{Stratton}
\begin{subequations}\label{poth}
	\begin{align}
	\vec{E}&=c_{TE}\vec{M}(\vec{r}) +  c_{TM} \vec{N}(\vec{r}),  \label{ec5}  \\
	\vec{H}&=-i \sqrt{\frac{\varepsilon}{\mu}}\left[ c_{TE}\vec{N}(\vec{r}) + c_{TM} \vec{M}(\vec{r})\right]\label{ec6}, 
	\end{align}
\end{subequations}
being $\vec{M}(\vec{r})$ and $\vec{N}(\vec{r})$ vector fields defined by
\begin{equation}
\label{ec:VecM}
\vec{M}(\vec{r})= \nabla \times [\hat{a} \psi(\vec{r})]
\end{equation}
and
\begin{equation}\label{ec:VecN}
\vec{N}(\vec{r})= \frac{1}{k} \nabla \times \vec{M}(\vec{r}),
\end{equation}
where  $\hat{a}$ is an arbitrary unit vector, $c_{TE}$ and $c_{TM}$ are two arbitrary complex numbers (the TE and TM sub indexes will be justified below), and $\psi(\vec{r})$ is a scalar field.\\
It is straightforward to verify that if the scalar field $\psi(\vec{r})$ satisfies the scalar Helmholtz equation,
\begin{equation}
\nabla^{2} \psi + k^{2} \psi =0,
\end{equation} 
then the fields (\ref{ec5}) and (\ref{ec6}) satisfy the vector Helmholtz equation. So, the scalar field $\psi(\vec{r})$ will be  named scalar potential. Note that these new vector fields, $\vec{M}$ and $\vec{N}$, are orthogonal between them, that is $\vec{M} \cdot \vec{N}=0$, and solenoidal, i.e. $\nabla \cdot \vec{M}=0$ and $\nabla \cdot \vec{N}=0 $.\\

\begin{figure}[htbp!]
\centering
\subfigure[Cylindrical coordinates: A Bessel beam with $\nu=3$.]{\includegraphics [scale=0.75]{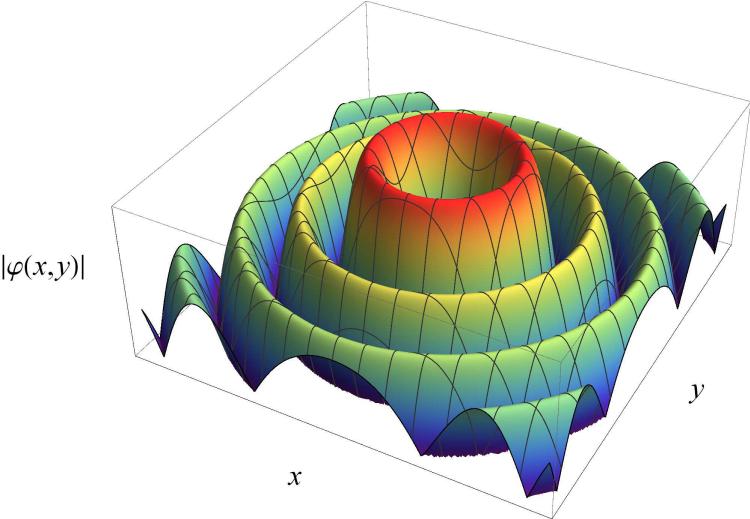}}
\qquad
\subfigure{\includegraphics [scale=0.50]{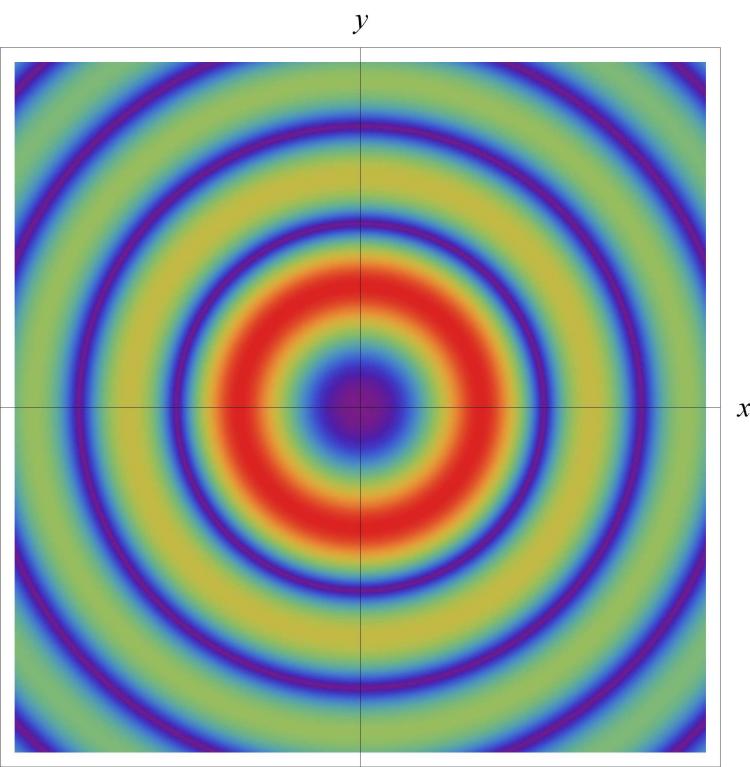}}
\subfigure[Parabolic cylindrical coordinates: An even Weber beam with $a=0$.]{\includegraphics[scale=0.75]{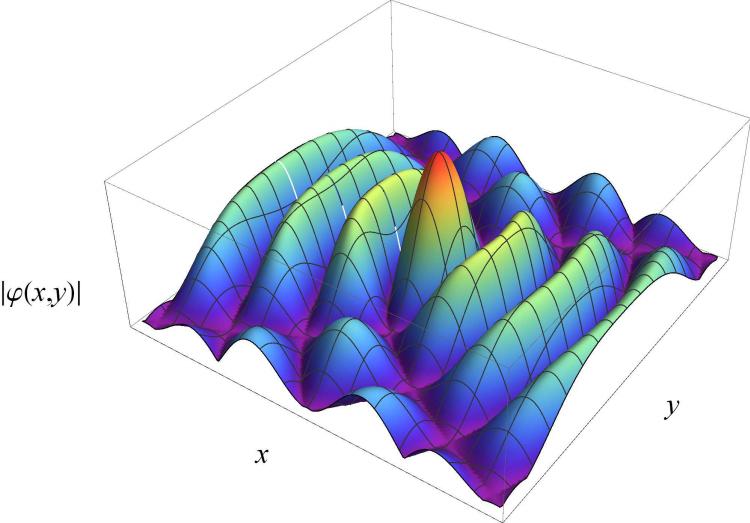}}
	\qquad
	\subfigure{\includegraphics[scale=0.50]{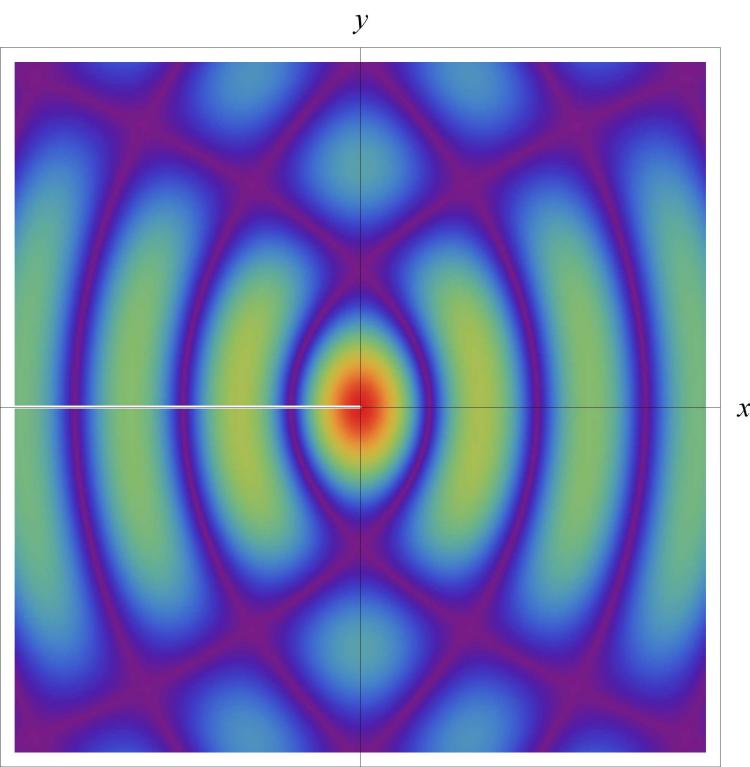}}
	\subfigure[Elliptical cylindrical coordinates: An even Mathieu beam with $q=25$, $n=3$ and $f=2 \sqrt{q}=11.28$.]{\includegraphics [scale=0.75]{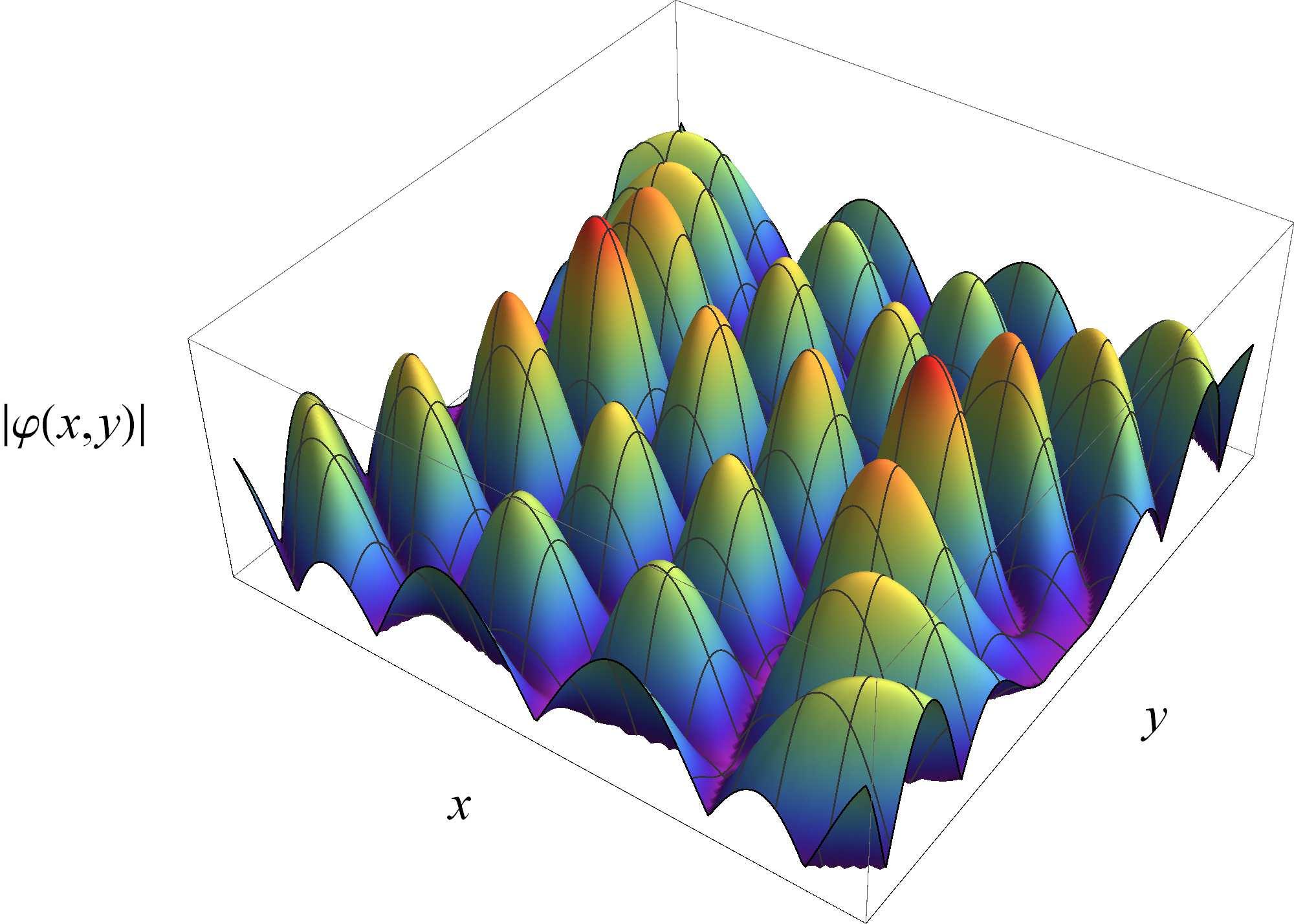}}
	\qquad
	\subfigure{\includegraphics [scale=0.50]{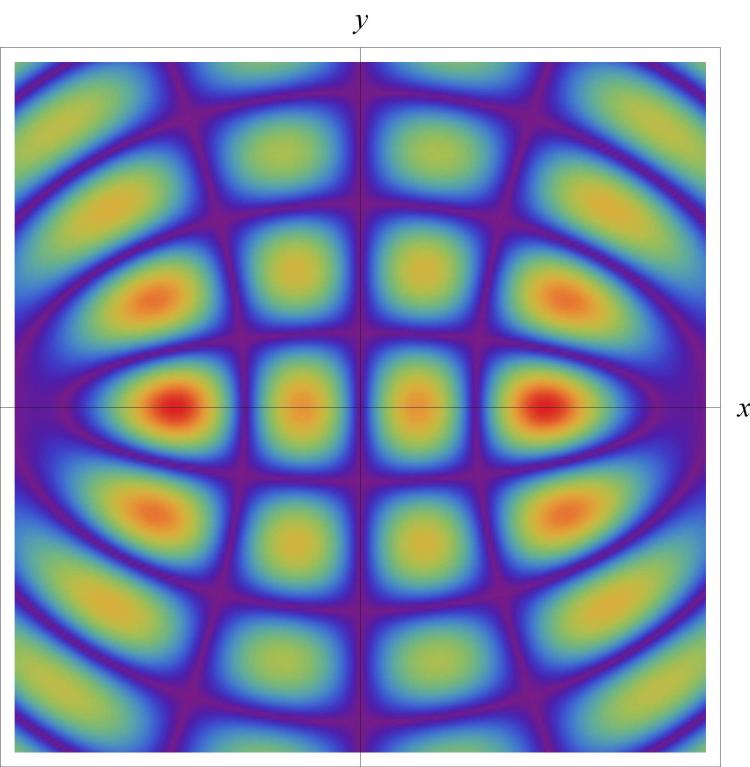}}
	\caption{Absolute value of the solution of the transversal Hemholtz equation (\ref{ec:HemholtzTrans}). In all graphics, we have used $k_T=1 \; \mathrm{m}^{-1}$. The scalar potential $\varphi$ is given in volts.}
	\label{Fig1}
\end{figure}

Though the homogeneous (source-free) Helmholtz equation can be separated in eleven coordinate systems, we require separability into transverse and longitudinal parts and that is possible only in Cartesian, cylindrical, parabolic cylindrical and elliptical cylindrical coordinates \cite{Boyer}. The spatial evolution of  the scalar potential $\psi$  can then be described by the transverse and the longitudinal parts; the transverse part $\varphi(u_1, u_2)$ will depend only on the transverse coordinates, $u_1,u_2$, and the longitudinal part $Z(z)$ will depend on the longitudinal coordinate $z$; thus
\begin{equation}
\label{ec:VarSepa}
\psi(u,v,z)=\varphi(u_1,u_2)Z(z).
\end{equation}
After  substituting (\ref{ec:VarSepa}) in the Helmholtz equation, we easily obtain that $\varphi(u_1, u_2)$ satisfy the two dimensional transverse Helmholtz equation
\begin{equation} \label{ec:HemholtzTrans}
\nabla _T^2\varphi + k^2_T \varphi=0,
\end{equation}
where $\nabla _T^2$ is the Laplacian transversal operator, which have a specific form in each coordinate systems, and the longitudinal part is $Z(z)=e^{i k_z z}$, with the dispersion relation $k^2= k_T^2 + k_z^2$.  In Figure (\ref{Fig1}), we show the transversal field distribution, given by the solution of (\ref{ec:HemholtzTrans}), for cylindrical, parabolic cylindrical and elliptical cylindrical coordinates.
To enhance the knowledge on these fields, we refer the reader to \cite{Bouchal,Sabino,Mariscal} for a general
description, physical properties, experiments and applications,  and for recent
advances to \cite{Hugo}.\\

We choose the unit vector $ \hat{a} $, in equation \eqref{ec:VecM}, as the unit vector that determines the direction of propagation, i.e. the $Z$ axis, and we note that in the four coordinate systems we are studying the scale factor $ h_3$ is equal to 1, to write
\begin{equation}\label{ec:OpM}
\vec{M}=-e^{i k_z z}\nabla _T^{\bot }\varphi,
\end{equation}
where
\begin{equation}\label{nablatp}
\nabla _T^{\bot }=-\hat{e}_1\frac{1}{h_2}\frac{\partial }{\partial u_2}+\hat{e}_2\frac{1}{h_1}\frac{\partial }{\partial u_1},
\end{equation}
$ \hat{e}_1 $ and $\hat{e}_2$ are the base unit vectors corresponding to the transversal direction, and $h_1$ and $h_2$ are the corresponding scale factors. It is also very easy to see that
\begin{equation}
\label{ec:OpN}
\vec{N}= \frac{e^{i k_z z}}{k}\left(  i k_z \nabla _T+\hat{e}_3 k_T^2\right)\varphi,
\end{equation}
where
\begin{equation}\label{nablat}
\nabla _T=\hat{e}_1\frac{1}{h_1}\frac{\partial }{\partial u_1}+\hat{e}_2\frac{1}{h_2}\frac{\partial }{\partial u_2}.
\end{equation}

\subsubsection{Example Bessel TE and TM modes }
Since  their introduction in 1987 by Durnin \cite{Durnin}, the Bessel beams  have attracted considerable attention due to their properties of propagation invariance and self-reconstruction, and they have found a wide range of interesting applications; for a historical review see \cite{McGloin},
for experimental mode realizations \cite{Flores} and for the vectorial approach \cite{Mishra,Olivik, Yu}.  Thus, as an example, we consider the case of the Bessel beams, which correspond to write the scalar transversal Helmholtz equation \eqref{ec:HemholtzTrans} in cylindrical coordinates. In this instance, the solution is
\begin{equation}\label{bessel}
\varphi(r,\theta)=  J_\nu  \left( k_{T}  r\right)   e^{i \nu \theta},
\end{equation}
where $\nu$ is a non-negative integer and $J_\nu(\zeta)$ is a Bessel function of the first kind of order $\nu$.\\
Let us substitute \eqref{bessel} and set $c_{TM}=0, c_{TE}=1$ in equations \eqref{poth}, to find
\begin{equation}\label{eq:Ete}
\vec{E}^{TE} =\mathbf{Re}\bigg( \frac{1}{r}
\left\lbrace  i \nu J_\nu( k_T r)  \hat{e}_r +  \left[\nu J_{\nu}( k_T r) -k_T r J_{\nu-1}( k_T r) \right] \hat{e}_{\theta}\right\rbrace
e^{i \nu \theta} e^{i k_z z}\bigg) ,
\end{equation}
and
\begin{equation}\label{ec:Hte}
\vec{H}^{TE} =\mathbf{Re}\bigg(
\sqrt{\frac{\varepsilon}{\mu}}\frac{1}{2 k r }
\left\lbrace
r k_T k_z \left[J_{\nu -1}\left(r k_T\right)-J_{\nu +1}\left(r k_T\right)\right]\hat{e}_r
+2 i \nu k_z J_{\nu }\left(r k_T\right) \hat{e}_{\theta} -2 i r k_T^2 J_{\nu }\left(r k_T\right) \hat{e}_z\right\rbrace e^{i \nu \theta} e^{i k_z z}\bigg) ,
\end{equation}
being $\hat{e}_r ,\, \hat{e}_{\theta},\, \hat{e}_z$ the unit base vectors in cylindrical coordinates. Note that the electric field is transversal; that means that its component in the propagation direction, $Z$, is zero. This justify \textit{a posteriori} the notation, the TE as super index in the electromagnetic fields and as subindex in the constant $c_{TE}$.\\
We substitute again \eqref{bessel} in equations \eqref{poth}, but we make now $c_{TE}=0,c_{TM}=1$, to find
\begin{equation}\label{eq:Etm}
\vec{E}^{TM} =\mathbf{Re}\bigg(\frac{1}{ k r}
\left\lbrace 
i k_z \left[r k_T J_{\nu -1}\left(r k_T\right)-\nu  J_{\nu }\left(r k_T\right)\right]\hat{e}_r
-\nu  k_z J_{\nu }\left(r k_T\right)\hat{e}_{\theta}
+r k_T^2 J_{\nu }\left(r k_T\right)\hat{e}_z
\right\rbrace
e^{i \nu  \theta} e^{i k_z z}\bigg),
\end{equation}
and 
\begin{equation}\label{eq:Htm}
\vec{H}^{TM}  = \mathbf{Re}\bigg(
\sqrt{\frac{\varepsilon }{\mu }}\frac{1}{2 r}
\left\lbrace  
2  \nu J_{\nu }\left(r k_T\right) \hat{e}_r 
+ i  r k_T \left[J_{\nu -1}\left(r k_T\right)-J_{\nu +1}\left(r k_T\right)\right]\hat{e}_{\theta}
\right\rbrace e^{i \nu  \theta} e^{i k_z z}\bigg) .
\end{equation}
Note now that the magnetic field is transversal and that is the reason why we use TM as super index in the electromagnetic field and as subindex in the constant $c_{TM}$.\\
Thus, in general, when $c_{TE}=1$ and $c_{TM}=0$ in  equations \eqref{poth}, we will get a transverse electric wave, and when $c_{TE}=0$ and $c_{TM}=1$ in  equations \eqref{poth}, we will get a transverse magnetic wave. \\

\section{Electromagnetic energy density }\label{SectionEnergyDensity}
The time averaged electromagnetic energy density $ \left\langle \mathcal{U}\right\rangle   $ for an harmonic field  in  an isotropic, linear and homogeneous medium is given by \cite{Stratton,Jackson,Griffiths}
\begin{align}\label{ec:energy}
\left\langle \mathcal{U}\right\rangle  =\frac{1}{4}\mathbf{Re} \left( \varepsilon\vec{E}\cdot \vec{E}^* + \mu \vec{H}\cdot \vec{H}^* \right),
\end{align}
where the upper index $\ast$ stands for the complex conjugate. Inserting the vector fields $\vec{M}$   and $\vec{N}$ in this formula, we obtain

\begin{align}\label{densener2}
\left\langle \mathcal{U}\right\rangle  &=\left(\left|c_\text{TE} \right|^2 +\left|c_\text{TM} \right|^2 \right) 
\Big( \left\langle \mathcal{U}\right\rangle _{\text{tra}} +  \left\langle \mathcal{U}\right\rangle  _{\text{z}}  \Big)    
+ \left\langle \mathcal{U}\right\rangle  _{\text{int}}  
\end{align}
where 
\begin{align}
\left\langle \mathcal{U}\right\rangle  _{\text{tra}}
& = \frac{\varepsilon}{4}\left( 1 + \frac{k_z^2}{k^2} \right)
\mathbf{Re}\left(  \nabla _T\varphi \cdot \nabla _T\varphi^*\right) ,
\end{align}
\begin{align}
\left\langle \mathcal{U}\right\rangle  _{\text{z}}
& =\frac{\varepsilon}{4} \frac{k_T^4}{k^2} \varphi \varphi^\ast 
\end{align}
and
\begin{align}\label{enerinter}
\left\langle \mathcal{U}\right\rangle  _{\text{int}}=  \varepsilon   \frac{ k_z}{2k} \,
\mathbf{Re}\left[ 
i \left(c_{\text{TE}} \, c_{\text{TM}}^*+c_{\text{TE}}^* \, c_{\text{TM}} \right) 
\left( \nabla _T^{\bot }\varphi \cdot \nabla _T\varphi^* \right) \right] .
\end{align}
The first term in \eqref{densener2} is the energy of the transverse and  longitudinal fields and \eqref{enerinter} is the energy of the \textquotedblleft interference\textquotedblright  $\,$  between the electric transversal modes and the magnetic transversal modes. We observe that the transversal electric and the transversal magnetic fields have the same energy, but that there is an interference term, that in general, will be different from zero, as the author of Reference \cite{Novitsky} noted studying the Poynting vector. It is worth to note, that the expression reported here, is valid for any invariant field and that the time averaged energy density does not depend on the longitudinal variable $z$, as  must be.

\subsubsection{Energy density for  Bessel beams}
In the case of Bessel beams of order $\nu$, the transversal energy density is 
\begin{align}
\left\langle \mathcal{U}\right\rangle  _{\text{tra}}=& \frac{\varepsilon}{4k^2 r^2}
\bigg\{ 
J^2_{\nu }\left(r k_T\right) \left[2 \nu ^2 \left(k^2+k_z^2\right)+r^2 k_T^4\right]+r^2 k_T^2 \left(k^2+k_z^2\right) J^2_{\nu -1}\left(r k_T\right)
\\ \nonumber &
-2 \nu  r k_T \left(k^2+k_z^2\right) J_{\nu -1}\left(r k_T\right) J_{\nu }\left(r k_T\right)
\bigg\},
\end{align}
and the interference energy density is
\begin{align}
\left\langle \mathcal{U}\right\rangle  _{\text{int}}= 
\frac{\varepsilon k_z}{k r^2}
\text{Re}\left\lbrace 
\left( c_\text{TE}c_\text{TM}^\ast+c_\text{TE}^\ast c_\text{TM}\right)
\nu J_{\nu }\left(r k_T\right) \left[r k_T J_{\nu -1}\left(r k_T\right)-\nu  J_{\nu }\left(r k_T\right)\right]\right\rbrace .
\end{align}
When $\nu=0$, i.e. for a Bessel beam of zero order, the total energy density simplifies to  
\begin{equation}\label{eq:UBesselZero}
\left\langle \mathcal{U}\right\rangle  = \frac{\varepsilon k_T^2}{4k^2}	\left(\left|c_\text{TE} \right|^2 +\left|c_\text{TM} \right|^2 \right)  \left[ k_T^2 J_0^2\left(k_T r \right) + \left(k^2+k_z^2 \right) J_1^2\left(k_T r \right) \right],
\end{equation}
where physically  the first term is related to the longitudinal component and the second one is due  to the transversal fields; similar expressions were reported by   \cite{Mishra,Olivik,Yu}. We want to remark, that choosing an appropriate $c_\text{TE}$ and $c_\text{TM}$ can lead to an linearly polarized Bessel beam with total angular momentum equal to the order of the Bessel function as linearly polarized basis do not carry orbital angular momentum. In the same manner, if $c_\text{TE}$ and $c_\text{TM}$  are chosen to deliver a right or left circularly polarized basis, for a Bessel field of total orbital  angular momentum $\nu$, the Bessel functions will have an order $\nu-1$ and $\nu+1$ as circularly polarized basis carry orbital angular momentum plus one and orbital angular momentum minus one \cite{Karem1}. In this work, we decided to focus on TEM modes.\\

\begin{figure}[htbp!]
	\centering
	\subfigure{\includegraphics [scale=0.65]{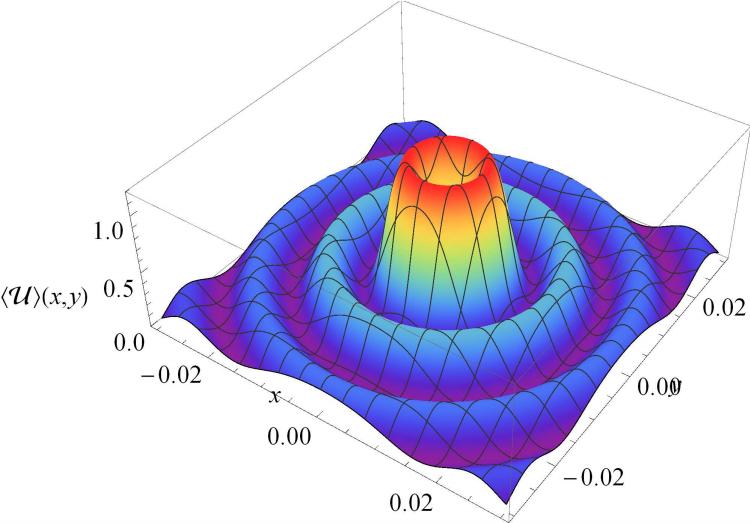}}
	\qquad
	\subfigure{\includegraphics [scale=0.65]{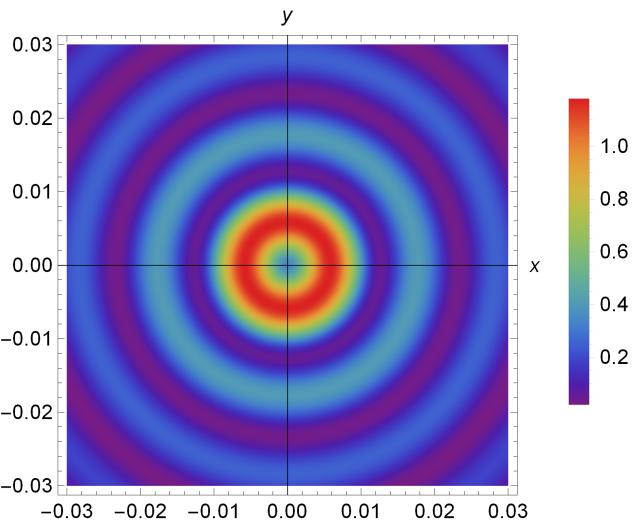}}
	\caption{The energy density, given by equation \eqref{eq:UBesselZero}, for a TE Bessel beam with $\nu=0$, $k=630 \; \mathrm{m}^{-1}$ and $k_T=300 \; \mathrm{m}^{-1}$.}
	\label{Fig2}
\end{figure}

Note that in the case of a Bessel beam of zero order the interference part is null. In Figure \eqref{Fig2}, we show  both contribution for the energy density for a Bessel beam  with  $\nu=0$. Remark also, that the Bessel beams have the energy distributed between its rings, so that the more rings they have, lower is the energy in the central core; this is important in many experimental applications (an interesting theoretical and experimental study was reported  in \cite{Lin}). \\

\section{Poynting vector} \label{SectionVectorPoynting}
For harmonic electromagnetic fields the time average power flow per unit area is given by \cite{Stratton,Jackson,Griffiths}
\begin{equation} \label{ec:VecPoyn}
\left\langle  \vec{S}\right\rangle  = \frac{1}{2} \mathbf{Re} \left(\vec{E} \times \vec{H}^* \right).
\end{equation}
Following the same procedure that in the energy density case and after performing the algebra, we obtain
\begin{align}
\left\langle  \vec{S} \right\rangle = \left|  c_{\text{TE}}\right| ^2\left\langle  \vec{S}_{\text{TE}} \right\rangle 
+\left|  c_{\text{TM}}\right| ^2 \left\langle  \vec{S}_{\text{TM}} \right\rangle 
+\left\langle \vec{S}_{\text{int}}\right\rangle ,
\end{align}
where
\begin{equation}
\left\langle \vec{S}_{\text{TE}}\right\rangle = \frac{1}{2 k }\sqrt{\frac{\varepsilon}{\mu}} \mathbf{Re}
\left[\left(\nabla_T\varphi \cdot \nabla_T\varphi^\ast\right)k_z \hat{e}_3  -i   k_T^2 \, \varphi^* \, \nabla _T\varphi\right] 
\end{equation}
is the transversal electric part,
\begin{equation}
\left\langle \vec{S}_{\text{TM}}\right\rangle = \frac{1}{2 k }\sqrt{\frac{\varepsilon}{\mu}} \mathbf{Re}
\left[\left(\nabla_T\varphi \cdot \nabla_T\varphi^\ast\right)k_z \hat{e}_3 +i   k_T^2 \, \varphi^* \, \nabla _T\varphi\right] 
\end{equation}
is the transversal magnetic part, and
\begin{align}
\left\langle \vec{S}_{\text{int}}\right\rangle =&  \frac{1}{2k^2} \sqrt{\frac{\varepsilon}{\mu}}  \mathbf{Re} \left[ i 
\left( c_{\text{TE}}\,c_{\text{TM}}^* k^2+   c_{\text{TE}}^*\,c_{\text{TM}}  k^2_z \right)   \left( \nabla _T^{\bot }\varphi \cdot \nabla _T\varphi^* \right) 
\hat{e}_3  
+ c_{\text{TE}}^*\,c_{\text{TM}}      k_z k_T^2  \,   \nabla_T^\bot  \left(\varphi \varphi^\ast \right) \right] 
\end{align}
is the interference part, which in general is not zero. Notice that this time averaged Poynting vector is independent of the $z$ coordinate, as expected.\\

In order to emphasizes the non-diffractive  character of these beams, it can be shown that the divergence of the transversal part of the time averaged Poynting vector is zero. To make this fact evident, let us write the time averaged Poynting vector as a sum of a transversal and a longitudinal part as
\begin{equation}
\left\langle  \vec{S} \right\rangle = \left\langle  \vec{S}_{\text{long}} \right\rangle 
+ \left\langle  \vec{S}_{\text{trans}} \right\rangle ,
\end{equation}
where
\begin{align}
\left\langle  \vec{S}_{\text{long}} \right\rangle =\hat{e}_3 \frac{1}{2k^2}\sqrt{\frac{\varepsilon}{\mu}}  \,  \mathbf{Re}
\left\lbrace 
\left( \left|c_{\text{TE}}\right| ^2 +\left|  c_{\text{TM}}\right| ^2\right)  k k_z
\left(\nabla_T \varphi \cdot \nabla_T \varphi^\ast\right) 
+i\left(c_{\text{TE}}c_{\text{TM}}^\ast k^2+c_{\text{TE}}^\ast c_{\text{TM}} k_z^2\right) \left( \nabla_T^\bot \varphi \cdot  \nabla_T \varphi^\ast  \right) 
\right\rbrace ,
\end{align}
and
\begin{align} \label{0310}
\left\langle  \vec{S}_{\text{trans}} \right\rangle = \frac{k_T^2}{2k^2} \sqrt{\frac{\varepsilon}{\mu}} \,
\mathbf{Re}\left[ 
-i \left( \left| c_{\text{TE}}\right| ^2    \varphi^\ast\nabla_T \varphi
-\left|  c_{\text{TM}}\right| ^2    \varphi\nabla_T \varphi^\ast    \right) k
+   c_{\text{TE}}^\ast  c_{\text{TM}}   k_z   \nabla_T^\bot\left(\varphi \varphi^\ast \right) 
\right] .
\end{align}
As we mention above, a diffraction free beam is such that the divergence of $ \left\langle  \vec{S}_{\text{trans}} \right\rangle$ is zero \cite{horak}. 
If we take the divergence of expression \eqref{0310}, we will get the terms $\nabla \cdot  \left( \varphi^\ast\nabla_T \varphi\right) $ and $\nabla \cdot  \left( \varphi\nabla_T \varphi^\ast \right)  $ which are equal, and as one is the complex conjugate of the other, they must be real and so  $\mathbf{Re}\left[ 
-i \left( \left| c_{\text{TE}}\right| ^2    \varphi^\ast\nabla_T \varphi
-\left|  c_{\text{TM}}\right| ^2    \varphi\nabla_T \varphi^\ast    \right) 
\right]=0$. Also $\nabla \cdot \nabla_T^\bot\left(\varphi \varphi^\ast \right) $ can be shown to be equal to zero, as $\nabla$ and $\nabla_T^\bot$ are orthogonal. Summarizing, $\nabla \cdot \left\langle  \vec{S}_{\text{trans}} \right\rangle =0 $ and so, following \cite{horak}, these beams are diffraction free; physically this means that the time averaged energy flux in the transverse direction is null.  Of course, it is possible to calculate explicitly the  transversal part of the Poynting vector in each of the four coordinate systems where we have separability into transversal and longitudinal parts (Cartesian, cylindrical, parabolic cylindrical and elliptic cylindrical) and take the divergence; that work had been done explicitly in the four cases and zero has been obtained.\\

\subsubsection{Poynting vector for Bessel beams}
\begin{figure}[htbp!]
	\centering
	\subfigure{\includegraphics [scale=0.6]{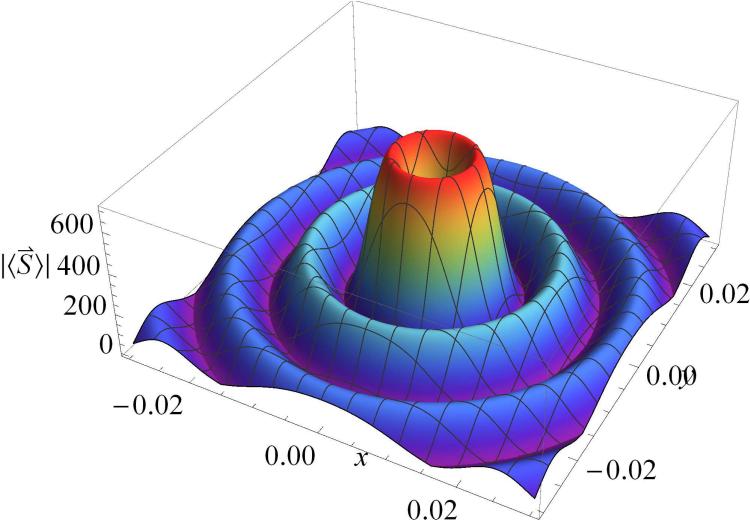}}
	\qquad
	\subfigure{\includegraphics [scale=0.47]{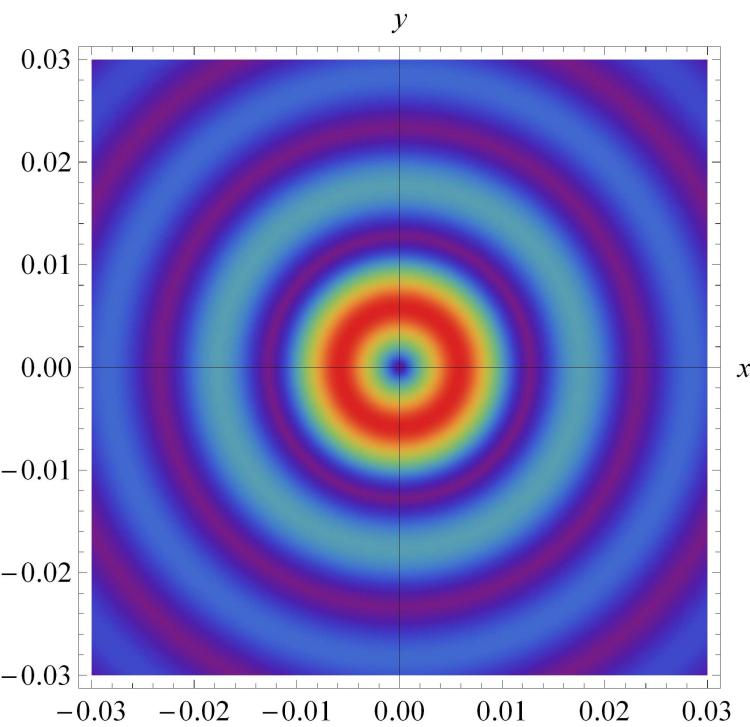}}
	\caption{The magnitude of the Poynting vector of a Bessel beam with $\nu=0$, $k=630 \; \mathrm{m}^{-1}$ and $k_T=300 \; \mathrm{m}^{-1}$. The beam is mixed, it has  $c_{\text{TE}}= c_{\text{TM}}=1$.}
	\label{Fig3}
\end{figure}

\begin{figure}[htbp!]
	\centering
	\subfigure{\includegraphics [scale=0.4]{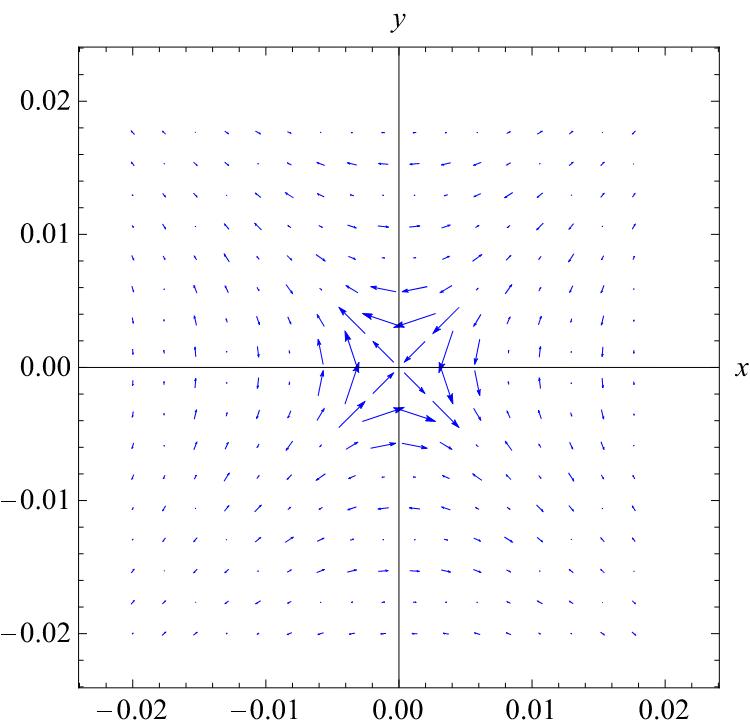}}
	\qquad
	\subfigure{\includegraphics [scale=0.4]{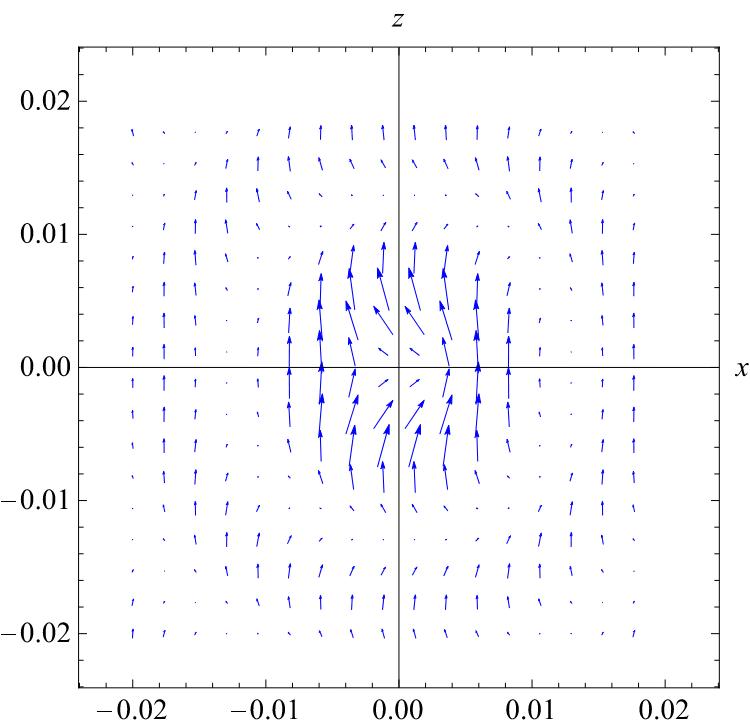}}
	\qquad
	\subfigure{\includegraphics [scale=0.4]{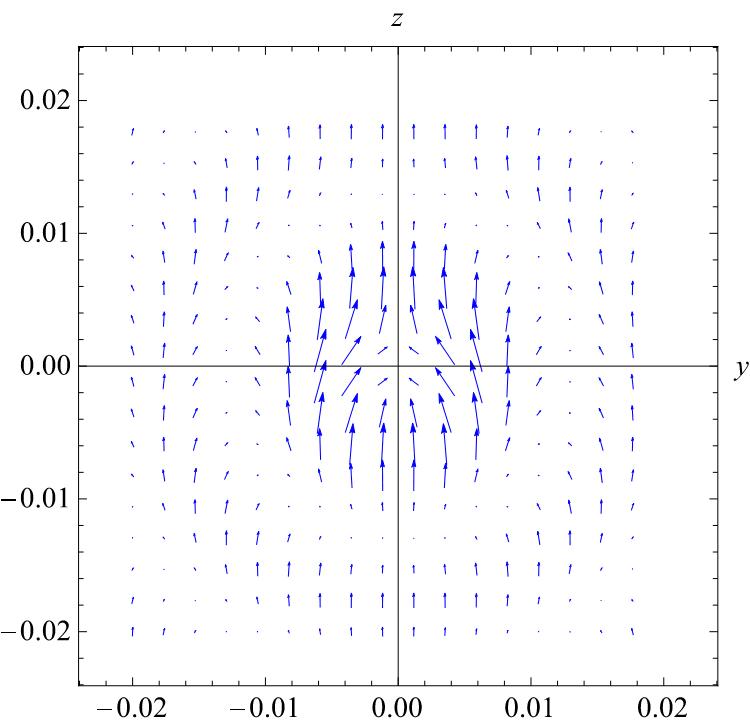}}
	\caption{Different projections of the Poynting vector of a Bessel beam of order zero with $k=630 \; \mathrm{m}^{-1}$ and $k_T=300 \; \mathrm{m}^{-1}$. The beam is mixed, it has  $c_{\text{TE}}= c_{\text{TM}}=1$.}
	\label{Fig4}
\end{figure}

We illustrate the previous results  calculating the Poynting vector  for  Bessel beams of order $\nu$; we get
\begin{subequations}
	\begin{align}
	\left\langle \vec{S}_r \right\rangle =& \sqrt{\frac{\varepsilon }{\mu }} \frac{k_T^2}{2kr}
	\mathbf{Re}
	\Big\{ i   
	\left( \left|  c_{\text{TE}}\right| ^2- \left|  c_{\text{TM}}\right| ^2\right) 
	J_{\nu }\left(r k_T\right) \left[ \nu  J_{\nu }\left(r k_T\right)-r k_T J_{\nu -1}\left(r k_T\right)\right]
	\Big\}=0, \\
	\left\langle \vec{S}_\theta \right\rangle =&
	\sqrt{\frac{\varepsilon }{\mu }} \frac{k_T^2}{2 k^2 r}
	\mathbf{Re} \Big(
	\left\lbrace 
	\left( \left|  c_{\text{TE}}\right| ^2+ \left|  c_{\text{TM}}\right| ^2\right) k \nu  J^2_{\nu }\left(r k_T\right)
	+2 c_{\text{TE}}^*\,c_{\text{TM}} k_z  J_{\nu }\left(r k_T\right)\left[ r k_T J_{\nu -1}\left(r k_T\right) - \nu J_{\nu }\left(r k_T\right)\right]
	\right\rbrace \Big),\\
	\left\langle \vec{S}_z\right\rangle  =&  
	\sqrt{\frac{\varepsilon }{\mu }} \frac{1}{2 k^2 r^2}
	\mathbf{Re} \Big(
	\Big\{
	\left( \left|  c_{\text{TE}}\right| ^2+ \left|  c_{\text{TM}}\right| ^2\right) k k_z 
	\left[
	r^2 k_T^2 J^2_{\nu -1}\left(r k_T\right)+2 \nu ^2 J^2_{\nu }\left(r k_T\right)-2 \nu  r k_T J_{\nu -1}\left(r k_T\right) J_{\nu }\left(r k_T\right)
	\right]
	\\ \nonumber 
	& +2 \left( k^2 c_{\text{TE}} \, c_{\text{TM}}^*  +k_z^2 c_{\text{TE}}^*\,c_{\text{TM}} \right)  
	\nu J_{\nu }\left(r k_T\right)
	\left[
	r k_T J_{\nu -1}\left(r k_T\right)-\nu  J_{\nu }\left(r k_T\right)
	\right] 
	\Big\}  \Big).
	\end{align}
\end{subequations}
The previous equations  resemble the  component expressions presented for $\vec{S}_{\theta}$ and $\vec{S}_z$  and the interference part  reported in \cite{Novitsky}. Note that in this case not only $\nabla \cdot \left\langle  \vec{S}_{\text{trans}} \right\rangle =0 $, but  $\left\langle \vec{S}_r \right\rangle  =0$; i.e., we have a zero transversal flux of energy \cite{horak}.\\
When $\nu=0$, we have a  zero order Bessel beam, and
\begin{subequations}\label{eq:BPoynting}
	\begin{align}
	\left\langle \vec{S}_r  \right\rangle =&
	\sqrt{\frac{\varepsilon }{\mu }}  \frac{k_T^3}{2 k}
	\mathbf{Re} \Big[
	i J_0\left(k_T r \right) J_1\left(k_T r \right)
	\left( 
	\left|  c_{\text{TE}}\right| ^2
	- \left|  c_{\text{TM}}\right| ^2
	\right) 	\Big]=0	 , \\
	\left\langle \vec{S}_\theta\right\rangle   =& 
	- \sqrt{\frac{\varepsilon }{\mu }}  \frac{k_z k_T^3}{k^2}
	J_0\left(k_T r \right) J_1\left(k_T r \right)
	\mathbf{Re} \left( 
	c_{\text{TE}}^* c_{\text{TM}}	\right) 	 , \\
	\left\langle \vec{S}_z  \right\rangle =& 
	\sqrt{\frac{\varepsilon }{\mu }}  \frac{k_z k_T^2}{2 k}
	J^2_1\left(k_T r \right)
	\left( 
	\left|  c_{\text{TE}}\right| ^2
	+ \left|  c_{\text{TM}}\right| ^2\right).
	\end{align}
\end{subequations}
We  show in Figure \eqref{Fig3} and \eqref{Fig4} the Poynting vector of a Bessel beam of zero order with $ c_{\text{TE}}= c_{\text{TM}}=1$.
Note that the Poynting vector transversal components circulate the beam center, as was reported theoretically \cite{Igor} and experimentally \cite{Mokhun}.\\

\section{Maxwell stress tensor}\label{SectionStressTensor}
Recently, the Maxwell stress tensor has  been used to calculate many optical properties of beams, such as the angular momentum \cite{Allen},  the density flux \cite{Barnett1994,Barnett} and the counter-propagation vortexes \cite{Mitri}. Some force and torque  problems  have also been approached using it \cite{Barton}. Also  the scattering of invariants beams for arbitrary  homogeneous dielectric particles (Bessel \cite{Marston1,Marston2}, Weber \cite{Chafiq}, Mathieu  \cite{Nebdi}) can be benefited of the employment of this tensor \cite{Zhiwei}.
However, a general formulation in terms of scalar fields has not been presented; such presentation can provide new physical insight, apart from simplifying the theoretical treatment by choosing only certain modes for a particular problem and also including \textquotedblleft interference between modes\textquotedblright.\\
The Maxwell tensor is a symmetric rank two tensor and it is useful to calculate the force interactions when using the Lorentz force is not a suitable alternative . It can be compared to the pressure tensor, where each $T_{ij}$ element can be interpreted as the force per unit of area parallel to the \textit{i}-th axis suffered by a surface normal to the \textit{j}-th  axis; the diagonal components represent the pressure while  the off-diagonal terms can be interpreted as  shear stress elements \cite{Griffiths}. The time average  stress tensor is given by \cite{Stratton,Jackson,Griffiths}
\begin{equation}
\overset{\leftrightarrow  }{ T } =\mathbf{Re}\left[  \frac{1}{2} \varepsilon \vec{E}\otimes\vec{E}^* +\frac{1}{2} \mu \vec{H}\otimes\vec{H}^* - \frac{\vec{E}\cdot\vec{E}^*+\vec{H}\cdot\vec{H}^* }{4} ( \hat{e}_1\otimes\hat e_1 + \hat e_2\otimes\hat e_2 + \hat e_3\otimes\hat e_3)\right] ,
\end{equation}
where $\otimes$  means the usual  outer product and $\hat {e}_i, \,  i=1,2,3$ are the basis orthonormal vectors. After a lengthy  calculation, we obtain the general expression for the Maxwell stress tensor,
\begin{align}
\overset{\leftrightarrow  }{ T } =
\left| c_{\text{TE}}\right| ^2 
\overset{\leftrightarrow  }{ T }_\text{TE}
+\left| c_{\text{TM}}\right| ^2  
\overset{\leftrightarrow  }{ T }_\text{TM}
+\overset{\leftrightarrow  }{ T }_{\text{int}} ,
\end{align}
where
\begin{align} 
\overset{\leftrightarrow  }{ T }_\text{TE} = &
\mathbf{Re}
\biggl\{ 
\frac{\varepsilon}{2k^2}
\left[ 
k_\text{T}^4 \varphi\varphi^*\hat{e}_3\otimes\hat{e}_3
+k^2 \gofi\otimes\gofi^*
+k_z^2   \gfi\otimes\gfi^*
+i k_z \ktc 
\left( \varphi^*\,\gfi\otimes\hat{e}_3
- \varphi\,\hat{e}_3\otimes\gfi^*\right) 
\right]
\nonumber   \\
& -\frac{1}{4}\bigg[ 
\frac{\varepsilon}{\mu}
\frac{1}{k^2} \left( 
k_\text{T}^4  \varphi\varphi^*
+k_z^2   \gfi\cdot\gfi^*
\right) 
+ \gofi\cdot\gofi^*
\bigg]  \overset{\leftrightarrow  }{ I } \biggr \} 
\end{align}
is the transversal electric stress tensor,
\begin{align}
\overset{\leftrightarrow  }{ T }_\text{TM} = &
\mathbf{Re}
\biggl\{ 
\frac{\varepsilon}{2k^2}
\left[ 
k_\text{T}^4 \varphi\varphi^*\hat{e}_3\otimes\hat{e}_3
+k^2 \gofi\otimes\gofi^*
+k_z^2   \gfi\otimes\gfi^*
+i k_z \ktc 
\left( \varphi^*\,\gfi\otimes\hat{e}_3
- \varphi\,\hat{e}_3\otimes\gfi^*\right) 
\right]
\nonumber   \\
& -\frac{1}{4}\bigg[ 
\frac{1}{k^2} \left( 
k_\text{T}^4  \varphi\varphi^*
+k_z^2   \gfi\cdot\gfi^*
\right) 
+\frac{\varepsilon}{\mu} \gofi\cdot\gofi^*
\bigg]  \overset{\leftrightarrow  }{ I }\biggr\} 
\end{align}
is the transversal magnetic stress tensor, and
\begin{align}
\overset{\leftrightarrow  }{ T }_{\text{int}} =&  \mathbf{Re}
\bigg\{ \Big( c_{\text{TE}}c_{\text{TM}}^*+c_{\text{TE}}^*c_{\text{TM}}\Big) 
\frac{\varepsilon}{2k}\bigg[  
- k_\text{T}^2 \left( \varphi\,\hat{e}_3\otimes\gofi^*
+ \varphi^*\gofi\otimes\hat{e}_3\right) 
+i k_z \left( \gofi\otimes\gfi^*
- \gfi\otimes\gofi^*\right) 
\bigg]  
\nonumber   \\
& \, - \frac{k_z}{4k} \bigg[ 
i \left( c_{\text{TE}}c_{\text{TM}}^*+\frac{\varepsilon}{\mu}c_{\text{TE}}^*c_{\text{TM}}\right)  
\gofi\cdot\gfi^*
-i\left(\frac{\varepsilon}{\mu} c_{\text{TE}}c_{\text{TM}}^*+c_{\text{TE}}^*c_{\text{TM}}\right)
\gfi\cdot\gofi^*
\bigg]  \overset{\leftrightarrow  }{ I } \bigg\} 
\end{align}
is the \textquotedblleft interference\textquotedblright  stress tensor.

\subsubsection{Maxwell stress tensor in cylindrical coordinates}
Using \eqref{bessel} in the expression we have obtained for the Maxwell stress tensor, we find the following components
\begin{align}
T_{1,1} =&  
\frac{\left|c_\text{TE} \right| ^2}{4 k^2 \mu  r^2} 
\Big\{
2 \nu ^2 J^2_{\nu }\left(r k_T\right)\big[(\varepsilon -1) k^2 \mu +\varepsilon  (\mu -1) k_z^2\big]
-\varepsilon  r^2  k_T^4 J^2_{\nu }\left(r k_T\right)
\nonumber \\
&+2 \nu  r k_T J_{\nu -1}\left(r k_T\right) J_{\nu }\left(r k_T\right)\big[k^2 \mu +(\varepsilon -2 \varepsilon  \mu ) k_z^2\big]
-r^2 k_T^2 J^2_{\nu -1}\left(r k_T\right)\big[k^2 \mu +(\varepsilon -2 \varepsilon  \mu ) k_z^2\big]
\Big\}
\nonumber \\
&+  \frac{\left|c_\text{TM} \right|^2}{4 k^2 \mu  r^2} 
\Big\{
2 \nu ^2 J^2_{\nu }\left(r k_T\right)\big[\varepsilon  k^2 (\mu -1)+(\varepsilon -1) \mu  k_z^2\big]
-\mu  r^2 k_T^4 J^2_{\nu }\left(r k_T\right)
\nonumber \\
&+2 \nu  r k_T J_{\nu -1}\left(r k_T\right) J_{\nu }\left(r k_T\right)\big[\varepsilon  k^2+(\mu -2 \varepsilon  \mu ) k_z^2\big]
-r^2 k_T^2 J^2_{\nu -1}\left(r k_T\right)\big[\varepsilon  k^2+(\mu -2 \varepsilon  \mu ) k_z^2\big]
\Big\}
\nonumber \\
&+\ctectm   \frac{k_z}{2 k \mu  r^2} 
\Big\{
\big[\varepsilon  (2 \mu -1)-\mu\big] \nu  J_{\nu }\left(r k_T\right)
\big[r k_T J_{\nu -1}\left(r k_T\right)-\nu  J_{\nu }\left(r k_T\right)\big]
\Big\},
\end{align}
\begin{align}
T_{2,2} =&
\frac{\left|c_\text{TE} \right| ^2}{4 k^2 \mu  r^2} 
\Big\{
2 \nu ^2 J^2_{\nu }\left(r k_T\right)\big[(\varepsilon -1) k^2 \mu +\varepsilon  (\mu -1) k_z^2\big]
-\varepsilon  r^2 k_T^4 J^2_{\nu }\left(r k_T\right)
\nonumber \\
&+ 2 \nu  r k_T J_{\nu -1}\left(r k_T\right) J_{\nu }\left(r k_T\right)\big[(1-2 \varepsilon ) k^2 \mu +\varepsilon  k_z^2\big]
+r^2 k_T^2 J^2_{\nu -1}\left(r k_T\right)\big[(2 \varepsilon -1) k^2 \mu -\varepsilon  k_z^2\big]
\Big\}
\nonumber \\
&+  \frac{\left|c_\text{TM} \right|^2}{4 k^2 \mu  r^2} 
\Big\{
2 \nu ^2 J^2_{\nu }\left(r k_T\right)\big[\varepsilon  k^2 (\mu -1)+(\varepsilon -1) \mu  k_z^2\big]
-\mu  r^2 k_T^4 J^2_{\nu }\left(r k_T\right)
\nonumber \\
&+2 \nu  r k_T J_{\nu -1}\left(r k_T\right) J_{\nu }\left(r k_T\right)\big[\varepsilon  k^2 (1-2 \mu )+\mu  k_z^2\big]
+r^2 k_T^2 J^2_{\nu -1}\left(r k_T\right)\big[\varepsilon  k^2 (2 \mu -1)-\mu  k_z^2\big]
\Big\}
\nonumber \\
&+\ctectm   \frac{k_z }{2 k \mu  r^2} 
\Big\{
\big[\varepsilon  (2 \mu -1)-\mu\big]\nu  J_{\nu }\left(r k_T\right)
\big[r k_T J_{\nu -1}\left(r k_T\right)-\nu  J_{\nu }\left(r k_T\right)\big]\Big\},
\end{align}
\begin{align}
T_{3,3}=&     \frac{
	\left|c_\text{TE} \right| ^2}{4 k^2 \mu  r^2} 
\Big\{
J^2_{\nu }\left(r k_T\right) \big[\varepsilon  (2 \mu -1) r^2 k_T^4-2 \nu ^2 \left(k^2 \mu +\varepsilon  k_z^2\right)\big]-r^2 k_T^2 \left(k^2 \mu +\varepsilon  k_z^2\right) J^2_{\nu -1}\left(r k_T\right)
\nonumber \\
& +2 \nu  r k_T \left(k^2 \mu +\varepsilon  k_z^2\right) J_{\nu -1}\left(r k_T\right) J_{\nu }\left(r k_T\right)
\Big\}
\nonumber \\
&+  \frac{\left|c_\text{TM} \right|^2}{4 k^2 \mu  r^2} 
\Big\{
J^2_{\nu }\left(r k_T\right) \big[(2 \varepsilon -1) \mu  r^2 k_T^4-2 \nu ^2 \left(\varepsilon  k^2+\mu  k_z^2\right)\big]-r^2 k_T^2 \left(\varepsilon  k^2+\mu  k_z^2\right) J^2_{\nu -1}\left(r k_T\right)
\nonumber \\
&+2 \nu  r k_T \left(\varepsilon  k^2+\mu  k_z^2\right) J_{\nu -1}\left(r k_T\right) J_{\nu }\left(r k_T\right)
\Big\}
\nonumber \\
& +\ctectm   \frac{k_z }{2 k \mu  r^2} 
\Big\{
\nu  (\varepsilon +\mu ) k_z J_{\nu }\left(r k_T\right) \big[\nu  J_{\nu }\left(r k_T\right)-r k_T J_{\nu -1}\left(r k_T\right)\big]
\Big\},
\end{align}
\begin{equation}
T_{1,2}=T_{2,1}=0,
\end{equation}
\begin{equation}
T_{1,3}= T_{3,1}=0,
\end{equation}
\begin{align}
T_{2,3}= T_{3,2}=& 
- \left( \left|c_\text{TE} \right| ^2 + \left|c_\text{TM} \right| ^2\right)   
\frac{k_T^2 k_z}{2 k^2  r} 
\varepsilon  \nu   J^2_{\nu }\left(r k_T\right)
\nonumber \\
&+ \ctectm  \frac{k_T^2}{2 k  r} 
\varepsilon   J_{\nu }\left(r k_T\right)
\big[\nu  J_{\nu }\left(r k_T\right)-r k_T J_{\nu -1}\left(r k_T\right)\big].
\end{align}

In the case of the zero order Bessel beam ($\nu=0$), we get
\begin{align}
\label{ec:Tcte11}
T_{1,1}=&
\left|c_\text{TE} \right| ^2   \frac{k_T^2}{4 k^2 \mu}
\Big\{
J^2_1\left(r k_T\right) \left[ \varepsilon  (2 \mu -1) k_z^2-k^2 \mu \right] -\varepsilon  k_T^2 J^2_0\left(r k_T\right)
\Big\}
\nonumber \\
& +\left|c_\text{TM} \right| ^2 \frac{k_T^2}{4 k^2 \mu}
\Big\{
-J^2_1\left(r k_T\right) \left[ \varepsilon  k^2+(1-2 \varepsilon ) \mu  k_z^2\right] -\mu  k_T^2 J^2_0\left(r k_T\right)
\Big\},
\end{align}
\begin{align}
\label{ec:Tcte22}
T_{2,2}=&
\left|c_\text{TE} \right| ^2   \frac{k_T^2}{4 k^2 \mu}
\Big\{
-J^2_1\left(r k_T\right)\left[(1-2 \varepsilon ) k^2 \mu +\varepsilon  k_z^2 \right] 
-\varepsilon  k_T^2 J^2_0\left(r k_T\right)
\Big\}
\nonumber \\
& +\left|c_\text{TM} \right| ^2  \frac{k_T^2}{4 k^2 \mu}
\Big\{
J^2_1\left(r k_T\right)\left[\varepsilon  k^2 (2 \mu -1)-\mu  k_z^2 \right] 
-\mu  k_T^2 J^2_0\left(r k_T\right)
\Big\},
\end{align}
\begin{align}
\label{ec:Tcte33}
T_{3,3}=&
\left|c_\text{TE} \right| ^2  \frac{k_T^2}{4 k^2 \mu}
\Big\{
\varepsilon  (2 \mu -1) k_T^2 J^2_0\left(r k_T\right)-J^2_1\left(r k_T\right) \left(k^2 \mu +\varepsilon  k_z^2\right)
\Big\}
\nonumber \\
&+\left|c_\text{TM} \right| ^2  \frac{k_T^2}{4 k^2 \mu}
\Big\{
(2 \varepsilon -1) \mu  k_T^2 J^2_0\left(r k_T\right)-J^2_1\left(r k_T\right) \left(\varepsilon  k^2+\mu  k_z^2\right)
\Big\},
\end{align}
\begin{align}
\label{ec:Tcte12and21}
T_{1,2}=& T_{2,1}=0,	\\	
\label{ec:Tcte13and31}		
T_{1,3}=& T_{3,1}=0,	\\	
\label{ec:Tcte23and32}			
T_{2,3}=& T_{3,2}=
\left( \cte\ctm^*+\cte^*\ctm\right) \frac{k_T^3}{2 k }
\varepsilon J_0\left(r k_T\right) J_1\left(r k_T\right).
\end{align}

\section{An example. The force over a cylinder}\label{cilindro}
In order to show and emphasize the utility of the Maxwell stress tensor, let us calculate the electromagnetic force  over a small cylinder on which a transversal electric ($c_{\mathrm{TM}}=0$) impinges a zero order Bessel field. The electromagnetic force is given in this case by \cite{Stratton,Jackson, Griffiths}
\begin{align}
\label{ec:ForceT}
F= \oint_S  \overset{\leftrightarrow  }{ T } \cdot \vec{da}.
\end{align}
The Maxwell stress tensor reduces to ($c_\text{TE}=1$ and $c_\text{TM}=0$)
\begin{align}
T_{1,1}=  \frac{k_T^2}{4 k^2 \mu}\Big\{
J^2_1\left(r k_T\right) \left[ \varepsilon  (2 \mu -1) k_z^2-k^2 \mu \right] -\varepsilon  k_T^2 J^2_0\left(r k_T\right)
\Big\},
\end{align}
\begin{align}
T_{2,2}=
\frac{k_T^2}{4 k^2 \mu}
\Big\{
-J^2_1\left(r k_T\right)\left[(1-2 \varepsilon ) k^2 \mu +\varepsilon  k_z^2 \right] 
-\varepsilon  k_T^2 J^2_0\left(r k_T\right)
\Big\},
\end{align}
\begin{align}
T_{3,3}=
\frac{k_T^2}{4 k^2 \mu}
\Big\{
\varepsilon  (2 \mu -1) k_T^2 J^2_0\left(r k_T\right)-J^2_1\left(r k_T\right) \left(k^2 \mu +\varepsilon  k_z^2\right)
\Big\},
\end{align}
\begin{align}
T_{1,2}= T_{2,1}=T_{1,3}=T_{3,1}=T_{2,3}= T_{3,2}=0.
\end{align}
We consider that the axis of the cylinder coincides with the $Z$ axis, has a longitude $2L$ and a radius $R$. The integrals over the two plane circular surfaces cancel each other and the integral over the curved side gives the following pressure in the radial direction
\begin{equation}
\mathcal{P}=\frac{k_T^2}{4k^2 \mu}\left\lbrace
\left[ \varepsilon\left( 2\mu-1\right)k_z^2-\mu k^2   \right] J_1^2\left( k_T R\right) 
-\varepsilon k_T^2  J_0^2\left( k_T R\right)
\right\rbrace .
\end{equation}
This pressure can be positive (directed outside) or negative (directed inside) depending on the parameters.

\section{Conclusions}\label{conclusiones}
We have shown how to obtain the principal properties for invariant propagation beams such as plane wave, Bessel, Mathieu  and Weber. Based in the scalar approach, we provide  general expressions for  the  energy density, the Poynting vector and the Maxwell stress tensor. In fact, these results can be used to study the orbital angular momentum of nonparaxial beams \cite{Brandao} and new optical phenomena, where these fields are present. Additionally, there are analytical expressions  to study the interaction between modes, as in \cite{Novitsky}. Furthermore, the scalar formalism can invite researchers to calculate electromagnetic properties  for any new optical field not discovered yet but written in a  scalar form  \cite{Boyer}. The present results should be of interest to a wide audience due to its fundamental character.

\section{Acknowledgment}
We acknowledge the ideas and supervision of B. M. Rodríguez-Lara during the first stage of research and are grateful for his continuous support and enlightening discussion. IRO acknowledges financial support from CONACYT graduate studies grant 423320.

\end{document}